\documentclass[12pt]{iopart}

\usepackage[dvips]{graphics,graphicx,color}

\begin{document}


\title[Forces within quantum Monte Carlo]{Methods for calculating
forces within quantum Monte Carlo simulations}

\author{A Badinski$^{1}$, P D Haynes$^{1,2}$, J R Trail$^{1,3}$, and R J
Needs$^1$}

\address{$^1$\ Theory of Condensed Matter Group, Cavendish Laboratory,
Cambridge CB3 0HE, UK}

\address{$^2$\ Department of Physics, Imperial College London,
Exhibition Road, London SW7 2AZ, UK}

\address{$^3$\ Japan Advanced Institute of Science and Technology,
School of Information Science, Asahidai 1-1, Nomi, Ishikawa 923-1292,
Japan}

\begin{abstract}
Atomic force calculations within the variational and diffusion quantum
Monte Carlo (VMC and DMC) methods are described.  The advantages of
calculating DMC forces with the ``pure'' rather than the ``mixed''
probability distribution are discussed.  An accurate and practical
method for calculating forces using the pure distribution is presented
and tested for the SiH molecule.  The statistics of force estimators
are explored and violations of the Central Limit Theorem are found in
some cases.
\end{abstract}

\submitto{\JPCM}
\maketitle

\section{Introduction \label{sec:introduction}}

Variational and diffusion quantum Monte Carlo (VMC and DMC) are
stochastic quantum Monte Carlo (QMC) methods for evaluating
expectation values with many-body wave functions \cite{foulkes_2001}.
The accuracy of VMC is quite limited in practice, but its more
accurate and sophisticated cousin the DMC method is capable of giving
very accurate results, often retrieving over 90\% of the correlation
energy.  The computational costs of fermion VMC and DMC calculations
scale approximately as the third power of the number of particles,
making it feasible to deal with hundreds or even thousands of
particles and allowing applications to electrons in condensed matter.

DMC has been applied to a wide variety of condensed matter systems,
including electron gases
\cite{ceperley_1980,drummond_2004,drummond_2008_2d}, nanocrystals
\cite{williamson_2002,drummond_2005_dia}, molecules on solid surfaces
\cite{filippi_2002,pozzo_2008a}, defects in semiconductors
\cite{leung_1999,hood_2003,alfe_2005a}, solid state structural phase
transitions \cite{alfe_2005b}, and equations of state
\cite{natoli_1993,delaney_2006,maezono_2007a,pozzo_2008b}.  DMC
calculations have provided accurate benchmark results for many
systems.  QMC algorithms are intrinsically parallel and are ideal
candidates for utilising the petascale computers
which are now becoming available.  Our \textsc{CASINO} code
\cite{casino} has already achieved efficient parallelisation on
machines with thousands of processors.

Calculating energy derivatives such as atomic forces is very important
in quantum mechanical simulations.  Forces are used in relaxing
structures, calculating their vibrational properties, and performing
molecular dynamics simulations.  It has, however, proved difficult to
develop accurate and efficient methods for calculating atomic forces
within DMC.  Difficulties have arisen in obtaining accurate
expressions which can readily be evaluated, and the statistical
properties of the force expressions are less advantageous than those
for the energy.  In this paper we study approaches for calculating
derivatives directly within QMC.  Finite-difference energy
calculations within QMC using correlated sampling methods
\cite{ceperley_1979} provide a different route to the same goal, and
such methods have been developed by other workers
\cite{filippi_2000,pierleoni_2005}.

\section{The VMC and DMC methods}
\label{sec:vmc and dmc}

In VMC the energy is calculated as the expectation value of the
Hamiltonian with an approximate many-body trial wave function
$\Psi_{\rm T}$.  For a real $\Psi_{\rm T}$ the energy is
\begin{eqnarray}
\label{eq:variational_energy}
E_{\rm V} = \frac{\int \Psi_{\rm T} ({\bf R}) \hat{H}({\bf R})
\Psi_{\rm T}({\bf R})\, d{\bf R}}{\int \Psi_{\rm T}({\bf R}) \Psi_{\rm
T}({\bf R})\, d{\bf R}} \;,
\end{eqnarray}
where $\hat{H}$ is the many-body Hamiltonian and ${\bf R}$ denotes the
vector of particle coordinates.  To facilitate stochastic evaluation,
$E_{\rm V}$ is written as
\begin{eqnarray}
\label{eq:variational_energy_2}
E_{\rm V} = \frac{\int \Psi_{\rm T}({\bf R})^2 E_{\rm L}({\bf R})\,
d{{\bf R}}}{\int \Psi_{\rm T}({\bf R})^2\, d{\bf R}} \;,
\end{eqnarray}
where the the local energy is
\begin{eqnarray}
\label{eq:local_energy}
E_{\rm L} = \Psi_{\rm T}^{-1} \hat{H} \Psi_{\rm T} \;.
\end{eqnarray}
The calculation proceeds by the generation of $N$ configurations ${\bf
R}_i$ sampled from the probability distribution
\begin{eqnarray}
\label{eq:vmc distribution}
P_{\rm V}({\bf R}) = \frac{\Psi_{\rm T}({\bf R})\Psi_{\rm T}({\bf R}) } 
{\int \Psi_{\rm T}({\bf R})  \Psi_{\rm T}({\bf R}) \, d{\bf R}} \;,
\end{eqnarray}
using, for example, the Metropolis algorithm, and the energy is
evaluated as
\begin{eqnarray}
\label{eq:variational_energy_3}
E_{\rm V} \simeq \frac{1}{N} \sum_{i=1}^N E_{\rm L}({\bf R}_i) \;.
\end{eqnarray}

Equation (\ref{eq:variational_energy_2}) with the $P_{\rm V}$ of
equation (\ref{eq:vmc distribution}) is an importance sampling
transformation of equation (\ref{eq:variational_energy}) which
exhibits the \textit{zero variance property}.  As $\Psi_{\rm T}$
approaches an exact eigenfunction, $E_{\rm L}$ becomes a smoother
function of ${\bf R}$ and the number of configurations, $N$, required
to achieve an accurate estimate of $E_{\rm V}$ is reduced.  If
$\Psi_{\rm T}$ is exact, equation~(\ref{eq:variational_energy_2})
gives the exact result even for a single configuration.  Although this
ideal limit cannot be reached in non-trivial calculations, it is
expected that using a zero-variance estimator will lead to improved
statistics.

DMC reduces the systematic bias inherent in VMC by an evolution of the
wave function in imaginary time.  Such projector methods suffer from
the infamous ``fermion sign problem'' in which the wave function
decays rapidly towards the lower energy bosonic ground state.  Stable
fermionic behaviour may, however, be achieved by using the
``fixed-node approximation'' \cite{anderson_1975,anderson_1976} in
which the nodal surface of the DMC wave function $\Phi$ is constrained
to equal that of $\Psi_{\rm T}$.
An importance sampling transformation is also used and the algorithm
generates configurations distributed according to the ``mixed''
probability distribution,
\begin{eqnarray}
\label{eq:dmc distribution}
P_{\rm D}({\bf R}) = \frac{\Psi_{\rm T}({\bf R}) \Phi({\bf R})} {\int
\Psi_{\rm T}({\bf R}) \Phi({\bf R})\, d{\bf R}}\;.
\end{eqnarray}
The DMC energy is given by
\begin{eqnarray} 
\label{eq:diffusion_energy 1}
E_{\rm D} = \frac{\int \Psi_{\rm T} \Phi \, E_{\rm L} \, d{\bf
R}}{\int \Psi_{\rm T} \Phi \, d{\bf R}} \;,
\end{eqnarray}
which is evaluated as 
\begin{eqnarray}
\label{eq:diffusion_energy 2}
E_{\rm D} \simeq \frac{1}{N} \sum_{i=1}^N E_{\rm L}({\bf R}_i) \;.
\end{eqnarray}
Both the VMC and DMC energies are upper bounds on the exact ground
state energy.  The DMC energy is, however, more accurate as it is
bounded from above by the VMC energy.

We also calculate expectation values with the pure probability
distribution,
\begin{eqnarray}
\label{eq:pure distribution}
P_{\rm P}({\bf R}) = \frac{\Phi({\bf R}) \Phi({\bf R})} 
{\int \Phi({\bf R})  \Phi({\bf R}) \,
d{\bf R}} \;.
\end{eqnarray}
Pure expectation values can be obtained using several methods: the
approximate (but often very accurate) extrapolation technique
\cite{ceperley_1979}, the future-walking technique \cite{barnett_1991}
which is formally exact but statistically badly behaved for large
systems and poor trial wave functions, and the reptation QMC technique
\cite{baroni_1999}, which is formally exact and well behaved, but
quite expensive.  The extrapolation technique can be used for any
operator, but the future-walking and reptation techniques are limited
to spatially local operators.

\section{Forces in VMC}
\label{sec:forces in vmc}

The derivative of the VMC energy of equation
(\ref{eq:variational_energy_2}) with respect to a parameter $\lambda$
in the Hamiltonian can be written as
\begin{eqnarray}
\label{eq:vmc force}
\frac{{\rm d}E_{\rm V}}{{\rm d}\lambda} 
= && \frac{ \displaystyle{\int \Psi_{\rm T}  \Psi_{\rm T} \, 
\left[ \frac{1}{\Psi_{\rm T}} 
\frac{{\rm d}\hat{H}}{{\rm d}\lambda}  \Psi_{\rm T} 
+ \frac{ (\hat{H} - E_{\rm L}) } {\Psi_{\rm T}}
\frac{{\rm d}\Psi_{\rm T}}{{\rm d}\lambda}
\right] \, d{\bf R}}} 
{\displaystyle{\int \Psi_{\rm T}  \Psi_{\rm T} \, d{\bf R}}} \nonumber 
\\ 
&& + 2 \frac{\displaystyle{\int \Psi_{\rm T}  \Psi_{\rm T} \,
\frac{ (E_{\rm L}-E_{\rm V}) }{\Psi_{\rm T}} \, 
\frac{{\rm d}\Psi_{\rm T}}{{\rm d}\lambda} \,
d{\bf R}}}{\displaystyle{\int \Psi_{\rm T}  \Psi_{\rm T} \, d{\bf R}}} \;,
\end{eqnarray}
This expression may be evaluated as an average over the probability
distribution $P_{\rm V}$ of equation (\ref{eq:vmc distribution}). The
term involving ${\rm d} \hat{H} / {\rm d}\lambda$ is the
Hellmann-Feynman force and the others are Pulay terms which contain
the derivative of $\Psi_{\rm T}$ with respect to the parameter in the
Hamiltonian. Calculating VMC forces including the Pulay terms is
straightforward, although it requires the evaluation of 
${\rm d}\Psi_{\rm T} / {\rm d} \lambda$.

Equation (\ref{eq:vmc force}) can be written in a number of different
ways which are equivalent when the expectation values are evaluated
exactly.  The particular form chosen satisfies the zero variance
condition if the configurations are sampled from $P_{\rm V}$ and
$\hat{H}$ and ${\rm d}\hat{H} / {\rm d}\lambda$ are taken to
act to the right.  If $\Psi_{\rm T}$ and ${\rm d}\Psi_{\rm
    T} / {\rm d}\lambda$ were exact, the exact force would be obtained
from averaging over any number of configurations.  It is expected that
this property will reduce the statistical noise in practical
calculations.  Equation (\ref{eq:vmc force}) has been used by a number
of groups to evaluate forces within VMC
\cite{assaraf_1999,casalegno_2003,assaraf_2003,lee_2005,badinski_2007,attaccalite_2008,badinski_2008c}.

\section{Forces in DMC}
\label{sec:forces in dmc}


The derivative of the DMC energy of equation (\ref{eq:diffusion_energy
1}) can be written as
\begin{eqnarray}
\label{eq:mixed dmc force}
\!\!\!\!\!\!\!\!\!\!\!\!\!\!\!\!\!\!\!\!\!\!\!\!   \frac{{\rm d}E_{\rm D}}{{\rm d} \lambda}
= \frac{
\displaystyle{ \int \Psi_{\rm T} 
\frac{{\rm d}\hat{H}}{{\rm d}\lambda} 
\Phi \,
d{\bf R}}}
{{\displaystyle \int \Psi_{\rm T} \Phi \, d{\bf R}}} 
+ \frac{ {\displaystyle \int 
\frac{{\rm d}\Psi_{\rm T}}{{\rm d}\lambda}
 (\hat{H}
-E_{\rm D}) \Phi \, d{\bf R}}}
{\displaystyle{\int \Psi_{\rm T} \Phi \, d{\bf R}}} 
+ \frac{{
\displaystyle {\int \Psi_{\rm T} (\hat{H}-E_{\rm D})
\frac{{\rm d}\Phi}{{\rm d}\lambda} 
\, d{\bf R}}}}
{{\displaystyle \int \Psi_{\rm T} \Phi \, d{\bf R}}} \;.
\end{eqnarray}
This expression includes the derivative of the DMC wave function ${\rm
  d}\Phi / {\rm d}\lambda$ which cannot be evaluated within the
standard approach.  A number of studies have used
equation~(\ref{eq:mixed dmc force}) with the approximation
\cite{reynolds_1986b}
\begin{eqnarray}
\label{eq:reynolds approximation} 
\frac{1}{\Phi}\frac{{\rm d}\Phi}{{\rm d}\lambda} 
\simeq 
\frac{1}{\Psi_{\rm T}} \frac{{\rm d}\Psi_{\rm T}}{{\rm d}\lambda} \;,
\end{eqnarray}
which leads to an expression which is no more difficult to evaluate
than the VMC one of equation (\ref{eq:vmc force}).  This approach
gives reasonable results if $\Psi_{\rm T}$ and 
${\rm d}\Psi_{\rm T} / {\rm d} \lambda$
are accurate enough, but it leads to errors of first order in
$(\Psi_{\rm T}-\Phi)$ and 
$({\rm d}\Psi_{\rm T} / {\rm d}\lambda
-{\rm d} \Phi / {\rm d}\lambda) $,
which are often significant in practice.

An alternative approach is to evaluate the forces using the pure
distribution $P_{\rm P}$ of equation (\ref{eq:pure distribution}),
which is proportional to $\Phi \Phi$.  It is more expensive to generate
configurations distributed according to the pure distribution than the
mixed one, but it brings significant advantages because less severe
approximations are required in evaluating the forces.  The energies
evaluated with the mixed and pure distributions are equal, i.e.,
\begin{eqnarray}
\label{eq:pure dmc energy}
E_{\rm D} 
= \frac{ 
\int \Psi_{\rm T} \hat{H} \Phi \, d{\bf R}}
{ \int \Psi_{\rm T} \Phi \, d{\bf R}} 
= \frac{  
\int \Phi \hat{H} \Phi \, d{\bf R}}
{
\int \Phi \Phi \, d{\bf R}} \;,
\end{eqnarray}
so that the DMC force evaluated with the pure distribution is
\begin{eqnarray}
\label{eq:pure dmc force}
\frac{{\rm d}E_{\rm D}}{{\rm d}\lambda} 
= \frac{{
\displaystyle \int \Phi \frac{{\rm d}\hat{H}}{{\rm d}\lambda} 
\Phi \, d{\bf
R}}}
{{\displaystyle \int \Phi \Phi \, d{\bf R}}} 
+ 2\frac{{
\displaystyle
\int \Phi (\hat{H}-E_{\rm D})
\frac{{\rm d}\Phi}{{\rm d}\lambda} 
\, d{\bf R}}}
{{\displaystyle \int \Phi \Phi \, d{\bf R}}} \;.
\end{eqnarray}

It turns out that the Pulay term in this expression can be written as
an integral over the nodal surface of $\Psi_{\rm T}$
\cite{huang_2000,schautz_2000,badinski_2008a}.  This fact was first
pointed out by Huang \textit{et al} \cite{huang_2000}, and Schautz and
Flad \cite{schautz_2000} provided an exact expression for the Pulay
term,
\begin{eqnarray} 
\label{nodal term 1}
2\frac{{
\displaystyle
\int \Phi (\hat{H}-E_{\rm D}) 
\frac{{\rm d}\Phi}{{\rm d}\lambda}
\, d{\bf R}}} 
{ \displaystyle {\int \Phi \Phi \, d{\bf R}}} 
= 
- \frac{1}{2} \frac{{
\displaystyle
\int_{\bf S} |\nabla\Phi| 
\frac{{\rm d}\Phi}{{\rm d}\lambda} 
\, d{\bf S}}}
{{
\displaystyle
\int \Phi \Phi \, d{\bf R}}} \;,
\end{eqnarray}
where ${\bf S}$ denotes the nodal surface and $\nabla\Phi$ is the
gradient of $\Phi$ obtained on approaching the nodal surface from
inside the nodal pocket along the direction normal to the surface.
Unfortunately this expression cannot readily be evaluated because it
involves ($i$) an integral over the nodal surface and ($ii$) the
quantity ${\rm d}\Phi / {\rm d}\lambda$.

One of the achievements of Badinski \textit{et al}
\cite{badinski_2008a} was to show that 
${\rm d}\Phi / {\rm d}\lambda$ can be
eliminated from equation (\ref{nodal term 1}) because the nodal
surfaces of $\Phi$ and $\Psi_{\rm T}$ must be equal for all values of
the parameter in the Hamiltonian.  They obtained the result
\begin{eqnarray} 
\label{nodal term 2}
2\frac{{
\displaystyle
\int \Phi (\hat{H}-E_{\rm D}) 
\frac{{\rm d}\Phi}{{\rm d}\lambda}
\, d{\bf R}}} 
{{
\displaystyle
\int \Phi \Phi \, d{\bf R}}} 
= - \frac{1}{2} \frac{{
\displaystyle
\int_{\bf S} \Phi \Phi \, 
\frac{|\nabla  \Psi_{\rm T}|}{\Psi_{\rm T} \Psi_{\rm T}}
\frac{{\rm d}\Psi_{\rm T}}{{\rm d}\lambda} 
\, d{\bf S}}}
{{
\displaystyle
\int \Phi \Phi \, d{\bf R}}} \;,
\end{eqnarray}
which is the average over the pure distribution of a quantity written
entirely in terms of $\Psi_{\rm T}$.  The difficulty of evaluating an
integral over the nodal surface remains and Badinski \textit{et al}
\cite{badinski_2008a,badinski_2008c} attempted to circumvent this
problem by writing the nodal surface integral as a volume integral.
They noted that the second term on the right hand side of equation
(\ref{eq:mixed dmc force}) can also be written as a nodal surface
integral, and that it is related to the pure distribution nodal term
of equations (\ref{nodal term 1}) and (\ref{nodal term 2}) by an
extrapolation approximation.  The standard extrapolation approximation
is
\begin{eqnarray} 
\label{extrapolation 1}
\frac{\int \Phi \hat{X} \Phi \, d{\bf R}} {\int \Phi \Phi \, d{\bf R}}
= 2 \frac{\int \Phi \hat{X} \Psi_{\rm T} \, d{\bf R}}{\int \Phi
\Psi_{\rm T} \, d{\bf R}} - \frac{\int \Psi_{\rm T} \hat{X} \Psi_{\rm
T} \, d{\bf R}}{\int \Psi_{\rm T} \Psi_{\rm T} \, d{\bf R}} +
{{\cal{O}}[(\Psi_{\rm T}-\Phi)^2}] \;,
\end{eqnarray}
where $\hat{X}$ is a Hermitian operator.  This expression can be
obtained by Taylor expansion in $(\Psi_{\rm T}-\Phi)$.

Application of equation (\ref{extrapolation 1}) to the nodal term of
equation (\ref{nodal term 2}) gives
\begin{eqnarray}
-\frac{1}{2} \frac{{
\displaystyle
\int_{\bf S} \Phi \Phi 
\frac{ |\nabla \Psi_{\rm T}|}{\Psi_{\rm T}\Psi_{\rm T}}
\frac{{\rm d}\Psi_{\rm T}}{{\rm d}\lambda} \, d{\bf S}}}
{{
\displaystyle
\int \Phi \Phi \,
  d{\bf R}}} 
= && 
-\frac{{
\displaystyle
\int_{\bf S} \Phi \Psi_{\rm T} \,
 \frac{|\nabla \Psi_{\rm T}|}{\Psi_{\rm T}\Psi_{\rm T}}
\frac{{\rm d}\Psi_{\rm T}}{{\rm d}\lambda} 
\,
  d{\bf S}}}
{\displaystyle{\int \Phi \Psi_{\rm T} \, d{\bf R}}} 
\nonumber 
\\ && +
  \frac{1}{2} 
  \frac{{
  \displaystyle
  \int_{\bf S} \Psi_{\rm T} \Psi_{\rm T} \,
  \frac{|\nabla \Psi_{\rm T}|}{\Psi_{\rm T}\Psi_{\rm T}}
  \frac{{\rm d}\Psi_{\rm T}}{{\rm d}\lambda} \,
  d{\bf S}}}
{
\displaystyle 
{\int \Psi_{\rm T} \Psi_{\rm T} \, d{\bf R}}} 
\nonumber
\\&& +  
{\cal{O}}[(\Psi_{\rm T}-\Phi)^2] \;.
\label{extrapolation 2}
\end{eqnarray}
It is straightforward to show that the variational term in equation
(\ref{extrapolation 2}) is zero because $\Psi_{\rm T}$ is continuous
and differentiable across the nodal surface \cite{badinski_2008a}.
The mixed-distribution term on the right hand side of equation
(\ref{extrapolation 2}) is, moreover, exactly equal to twice the
second term on the right hand side of equation (\ref{eq:mixed dmc
force}) \cite{badinski_2008a}, and therefore
\begin{eqnarray}
\label{extrapolation 3}
-\frac{1}{2} \frac{{
\displaystyle
\int_{\bf S} \Phi \Phi \, 
\frac{|\nabla  \Psi_{\rm T}|}{\Psi_{\rm T} \Psi_{\rm T}} 
\frac{{\rm d}\Psi_{\rm T}}{{\rm d}\lambda}
\, d{\bf S}}}
{{
\displaystyle
\int \Phi \Phi \,
  d{\bf R}}} 
= && 
2 \frac{{
\displaystyle
\int \Psi_{\rm T} \Phi \, 
\frac{(\hat{H}-E_{\rm D})}{\Psi_{\rm  T}} 
\frac{{\rm d}\Psi_{\rm T}}{{\rm d}\lambda}
 \, d{\bf R}}}
{{
\displaystyle
\int
  \Psi_{\rm T} \Phi \, d{\bf R}}} 
\nonumber 
\\ && +
  {\cal{O}}[(\Psi_{\rm T}-\Phi)^2] \;.
\end{eqnarray}
The mixed-distribution term on the right hand side of equation
(\ref{extrapolation 3}) may be evaluated straightforwardly if
$\hat{H}$ is taken to act to the right.  Next we apply the
extrapolation procedure again to eliminate the mixed distribution term
in equation (\ref{extrapolation 3}), which leads to
\begin{eqnarray}
\label{extrapolation 4}
\!\!\!\!\!\!\!\!\!\!\!\!\!\!\!\!\!\!\!\!\!\!\!\! -\frac{1}{2} \frac{{
\displaystyle
\int_{\bf S} \Phi \Phi \, 
\frac{|\nabla \Psi_{\rm T}|}{ \Psi_{\rm T}  \Psi_{\rm T}} 
\frac{{\rm d}\Psi_{\rm T}}{{\rm d}\lambda} 
\, d{\bf S}}}
{{
\displaystyle
\int \Phi \Phi \, d{\bf R}}} 
= && 
\frac{{
\displaystyle
\int \Phi \Phi \,
\frac { (\hat{H}-E_{\rm D}) }{ \Psi_{\rm T}}
\frac{{\rm d}\Psi_{\rm T}}{{\rm d}\lambda}
\, d{\bf R}}}
{{
\displaystyle
\int \Phi  \Phi \, d{\bf R}}} 
\nonumber 
\\ && + \frac{{
\displaystyle
\int \Psi_{\rm T} \Psi_{\rm T}  
\frac{(\hat{H}-E_{\rm D})}{\Psi_{\rm T}} 
\frac{{\rm d}\Psi_{\rm T}}{{\rm d}\lambda} 
\, d{\bf R}}}
{
\displaystyle
{\int \Psi_{\rm T} \Psi_{\rm T} \, d{\bf R}}}
+ {\cal{O}}[(\Psi_{\rm T}-\Phi)^2] \;.
\end{eqnarray}
Using equations (\ref{eq:pure dmc force}), (\ref{nodal term 2}), and
(\ref{extrapolation 4}) with $\hat{H}$ acting to the left in the
variational term, we obtain our final expression for the energy
derivative \cite{badinski_2008c}
\begin{eqnarray}
\frac{{\rm d}E_{\rm D}}{{\rm d}\lambda} 
= && \frac{{
\displaystyle
\int \Phi \Phi \, \left[ 
\frac{1}{\Phi} 
\frac{{\rm d}\hat{H}}{{\rm d}\lambda}
\Phi
+ 
\frac{(\hat{H}-E_{\rm D})}{\Psi_{\rm T}} 
\frac{{\rm d}\Psi_{\rm T}}{{\rm d}\lambda} 
\right] \, d{\bf R}}}
{{
\displaystyle
\int \Phi \Phi \, d{\bf R}}} \nonumber
\\&&
+ \frac{{
\displaystyle
\int \Psi_{\rm T} \Psi_{\rm T} \,
\frac{(E_{\rm L}-E_{\rm D})}{\Psi_{\rm T}}
\frac{{\rm d}\Psi_{\rm T}}{{\rm d}\lambda}
\, d{\bf
R}}}
{{
\displaystyle
\int \Psi_{\rm T} \Psi_{\rm T} \, d{\bf R}}}  
+
{\cal{O}}[(\Psi_{\rm T}-\Phi)^2] \;.
\label{final pure force}
\end{eqnarray}
Equation (\ref{final pure force}) consists of integrals over the pure
and variational distributions.  This expression is expected to give
much more accurate forces than equation (\ref{eq:mixed dmc force})
with the approximation (\ref{eq:reynolds approximation}) because ($i$)
the Pulay terms with the pure distribution are smaller than those
obtained with the mixed distribution, and ($ii$) the error is of
second order rather than first order.  Equation (\ref{final pure
force}) is expected to show reasonable statistical behaviour because
it satisfies a zero-variance principle.  If $\Psi_{\rm T}$ and
${\rm d}\Psi_{\rm T} / {\rm d}\lambda$ 
are exact, the pure distribution term in
equation (\ref{final pure force}) gives the exact result even for a
single configuration, and the variational term has the same property
and, in addition, gives zero if $\Psi_{\rm T}$ is exact.

If the nodal surface of $\Psi_{\rm T}$ is independent of the variable
parameter in the Hamiltonian then only the Hellmann-Feynman term
survives in equation (\ref{final pure force}).  The fixed-node
approximation is equivalent to placing infinite potential barriers
everywhere on the nodal surface of $\Psi_{\rm T}$.  The nodal surface
term can be viewed as arising from the change in the potential
barriers as the variable parameter changes.  Alternatively, one can
include the infinite potential barriers in the Hamiltonian,
formulating a fixed-node Hamiltonian $\hat{H}_{\rm FN}$
\cite{badinski_2008a}.  The Hellmann-Feynman theorem holds for
$\hat{H}_{\rm FN}$, but $\hat{H}_{\rm FN}$ depends on the nodal
surface of $\Psi_{\rm T}$ and the derivative 
${\rm d}\hat{H}_{\rm FN} / {\rm d}\lambda$ 
generates the same Pulay terms as in the approach
described above.

\section{Total versus partial derivatives of $\Psi_{\rm T}$}
\label{forces:total_partial}

%
%

We assume that the trial wave function $\Psi_{\mathrm{T}}$ (and hence
the nodal surface) depends on $N_{\mathrm{c}}$ variational parameters
$c_{i}$, with $i=1,...,N_{\mathrm{c}}$.  We further assume that
$\Psi_{\mathrm{T}}$ has both explicit and implicit dependence on the
parameter $\lambda$ in the Hamiltonian we wish to vary.  For example,
the atomic centres of a localised basis set or the Jastrow factor may
have explicit dependence on the positions of the atoms, and they may
contain variable parameters whose optimal values depend implicitly on
the atomic positions.  The total force calculated with the
variational, mixed or pure distributions can be written as
\begin{eqnarray}
\frac{{\rm d}E}{{\rm d}\lambda} \biggr|_{\lambda =0}
=\frac{\partial E}{\partial\lambda}\biggr|_{\lambda =0}
+\sum_{i=1}^{N_{\mathrm{c}}}\frac{\partial E}{\partial c_{i}}
\frac{{\rm d}c_{i}}{{\rm d}\lambda} \biggr|_{\lambda =0},
\label{eq:F:F_total_partial}
\end{eqnarray}
where $E \equiv E_{\rm V}$ for VMC and $E \equiv E_{\rm D}$ for the
mixed and pure DMC distributions.  The two terms on the right hand
side of equation (\ref{eq:F:F_total_partial}) correspond to the
explicit and implicit dependences on $\lambda$. The choice of these
dependences is not unique and may depend on how the parameters are
considered in a QMC simulation. It is possible to construct
$\Psi_{\mathrm{T}}$ without variational parameters if the values of
all of the parameters $c_{i}$ are kept fixed.  Also,
$\Psi_{\mathrm{T}}$ can be constructed so that it has no explicit
dependence on the nuclear coordinates when, for example, the centres
of atomic-centred basis sets are made variational
parameters~\cite{casalegno_2003}. For QMC calculations it is, however,
convenient to choose forms of $\Psi_{\mathrm{T}}$ where both explicit
and implicit dependences on $\lambda$ are present.

Calculating the force using equation (\ref{eq:F:F_total_partial})
involves knowledge of how the wave function and hence the $c_{i}$
change with $\lambda$, i.e., $ {\rm d} c_i / {\rm d} \lambda$. In
standard quantum chemistry methods, these derivatives are typically
obtained by second-order perturbation
theory~\cite{takada_1983}. Unfortunately such an approach is not
straightforward in VMC and DMC.

We follow a different route and approximate all total derivatives by
partial derivatives or, equivalently, the term involving the sum in
equation (\ref{eq:F:F_total_partial}) is neglected.  This
approximation introduces an error in the force of first order in
$(\Psi_{\mathrm{T}}-\Phi)$. We expect this approximation to be rather
accurate in both VMC and DMC for the following reasons.  In VMC each
term in the sum vanishes if $E$ is energy-minimized with respect to
the corresponding $c_{i}$, i.e., $\partial E / \partial c_i
=0$~\cite{casalegno_2003}.  In practice this condition will not
normally be satisfied exactly because $i)$ the $c_{i}$ are obtained by
stochastic methods and are therefore subject to statistical noise,
$ii)$ the $c_{i}$ are sometimes obtained by minimizing the variance of
the local energy rather than the energy itself, $iii)$ the orbital
parameters are often obtained from density-functional theory
calculations and are therefore energy minimised with respect to the
wrong Hamiltonian.  Similar considerations apply within DMC, and we
expect that $E$ is approximately minimized with respect to the $c_{i}$
so that the $\partial E / \partial c_i$ are very small.  The terms in
the sum in equation (\ref{eq:F:F_total_partial}) are therefore
expected to be very small and can be neglected.

\section{Practical test of DMC forces}
\label{sec:practical dmc forces}

Before presenting some results, it is worth discussing the accuracy we
require of QMC forces for them to be useful.  It is helpful to think
about the error in the force on an atom in a molecule or solid in
terms of the associated error in the bond length.  The errors in
equilibrium bond lengths obtained from well-converged density
functional theory (DFT) calculations for $sp$-bonded atoms are
$\left(a_{\rm DFT} - a_{\rm exp}\right) \sim$ 0.01\,\AA, where ``exp''
denotes the experimental value.  To be useful, the errors in DMC bond
lengths measured as the difference between the values obtained from
the forces and energies $\left(a^{\rm Forces}_{\rm DMC} - a^{\rm
Energies}_{\rm DMC}\right)$ should be substantially smaller than
$\sim$0.01\,\AA.

We illustrate the results obtained for the mixed and pure DMC force
estimates with calculations on the SiH molecule.  It should be noted
that we have calculated the forces on both the Si and H atoms, which
should of course be of equal magnitude and opposite sign.  This is in
fact a highly non-trivial test of the computational method as the
condition of zero net force on the molecule is not automatically
satisfied in these calculations.  It is easier to obtain accurate
trial wave functions for H atoms than for heavier atoms and therefore
the forces on the H atoms tend to be more accurate.  A number of other
DMC studies have calculated forces on small molecules containing H
atoms in which the bond lengths were deduced entirely from the forces
on the H atoms.  We do not regard this procedure as a proper test of
the methodology.

In our calculations we used Dirac-Fock based non-local
pseudopotentials to represent the H$^+$ and Si$^{4+}$ ions
\cite{trail_2005_1,trail_2005_2,casino_page}.  The Hellmann-Feynman
contributions to the forces from the non-local pseudopotentials were
calculated using the expressions given in the Appendix of
Ref.\ \cite{badinski_2007}.  The pseudopotential localisation procedure
\cite{mitas_1991} introduces additional terms in the force which
contain ${\rm d}\Psi_{\rm T} / {\rm d}\lambda$ 
\cite{badinski_2008b}.  These terms
arise from the derivative of the Hamiltonian and therefore we have
included them in the HFT expression for the forces in the following
discussion.

The trial wave functions were constructed as follows.  We calculated
the molecular orbitals within Hartree-Fock theory using the GAMESS
code \cite{gamess} with a fairly small basis set consisting of 5$s$
and 2$p$ Gaussian basis functions for each atom.  We used a Jastrow
factor which included electron-electron and electron-ion terms
\cite{ndd_newjas}.  The initial values of the Jastrow parameters were
obtained by minimising the variance of the energy
\cite{drummond_2005_min} and they were further refined by minimising
the VMC energy \cite{umrigar_emin,toulouse_emin,brown_2007}.  We could
have generated considerably more accurate trial wave functions by
using a larger Gaussian basis set, a more complicated form of Jastrow
factor, including several determinants, or backflow transformations
\cite{lopez-rios_2006}.  We chose not to do this because we wanted to
work with trial wave functions of a similar quality to those we could
readily generate for heavier atoms and much larger systems.  When
evaluating ${\rm d}\Psi_{\rm T} / {\rm d} \lambda$ 
we included only the explicit
dependence on the ionic positions which occurs in the Jastrow factor
and Gaussian basis functions.  Expectation values with the pure
distribution were evaluated using the future-walking method
\cite{barnett_1991}.

\begin{figure}[t!]
\centering
\includegraphics*[width=.7\textwidth]{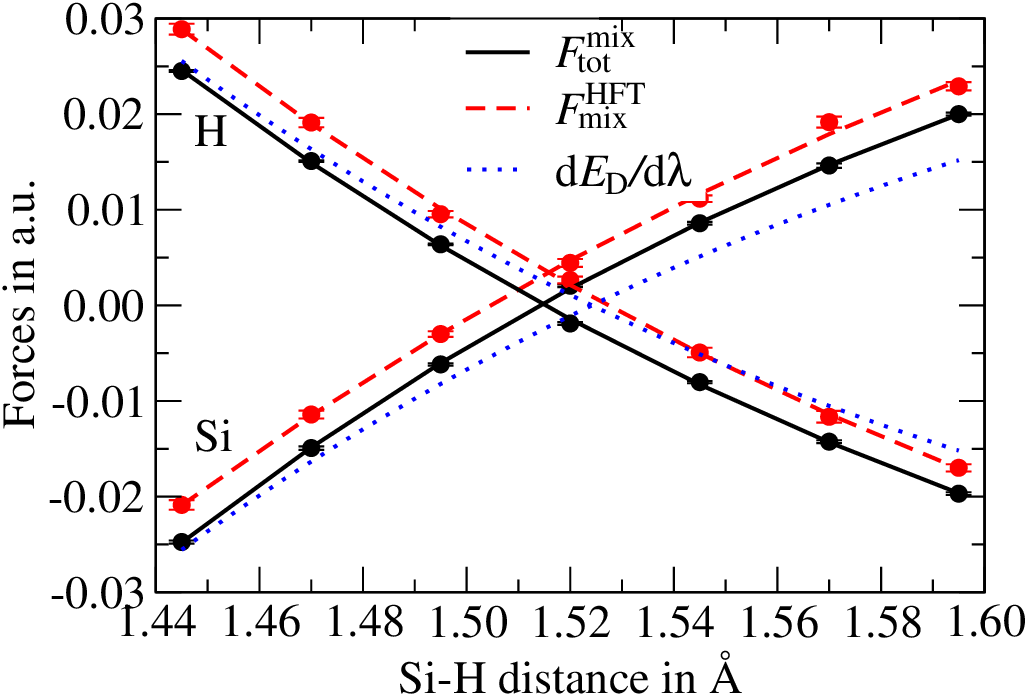}
\caption[]{
The mixed estimates of the Hellmann-Feynman force ($F_{\rm mix}^{\rm
  HFT}$) and the total force ($F_{\rm mix}^{\rm tot}$) evaluated for
the H and Si atoms. The derivative of the Morse potential fitted to
the DMC energies (${\rm d}E_{\rm D} / {\rm d}\lambda$) is shown as
reference.
}
\label{fig:mixed}   
\end{figure}

Results using the mixed distribution expression of equation
(\ref{eq:mixed dmc force}) with the approximation of (\ref{eq:reynolds
approximation}) are plotted in figure \ref{fig:mixed}. As reference,
we have also plotted the derivative of the Morse potential previously
obtained from a fit to the DMC energies.  The minimum in the DMC
energy occurs at a bond length of 1.5242(6)\,\AA. The forces on the H
and Si atoms should be equal and opposite, and figure \ref{fig:mixed}
shows that the total forces $F_{\rm mix}^{\rm tot}$ on the H and Si
atoms sum almost to zero over the range of bond lengths shown, with an
error corresponding to an error in the bond length of equal or less
than 0.001\,\AA.  However, the total force gives an equilibrium bond
length which is about 0.010\,\AA\ too short.  The Hellmann-Feynman
force $F_{\rm mix}^{\rm HFT}$ on the H atom is zero at a bond length
which is 0.003(1)\,\AA\ too long, while that on the Si atom is zero at
a bond length which is 0.020(1)\,\AA\ too short.
%
%
Including the Pulay terms improves the agreement between the forces calculated
on the H and Si atoms, but overall the bond lengths from the mixed-estimator
forces are no better than those which could be obtained in DFT calculations.

\begin{figure}[t!]
\centering
\includegraphics*[width=.7\textwidth]{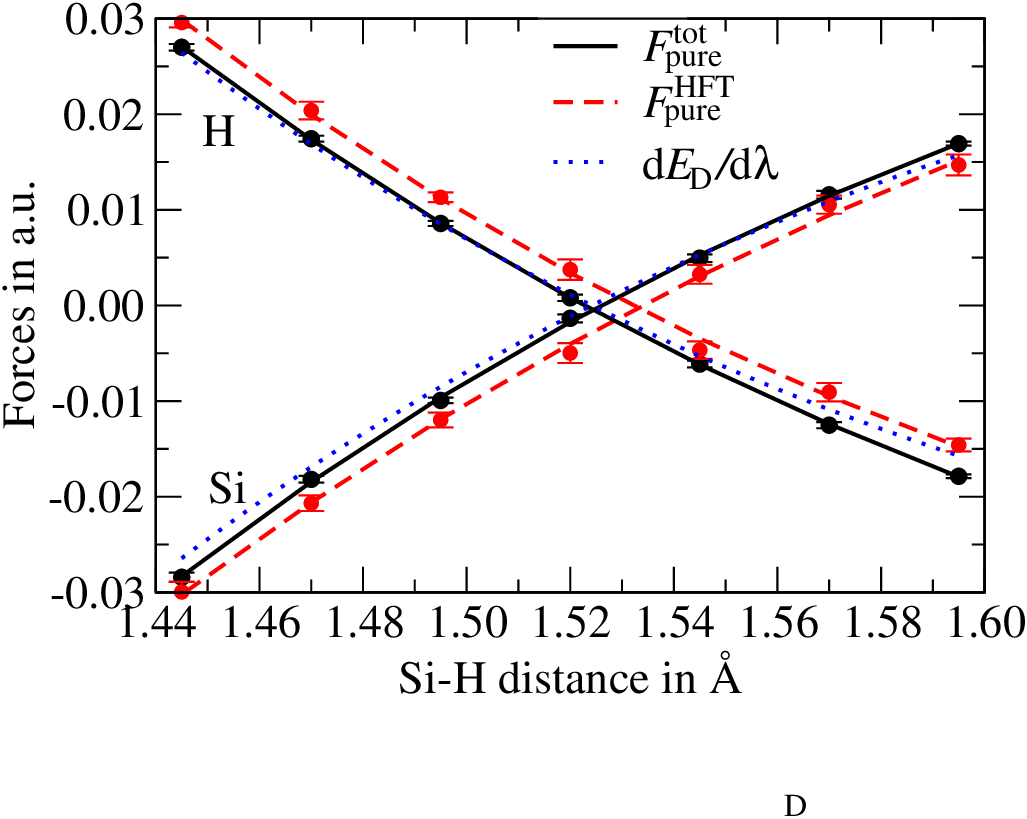}
\caption[]{The same as figure \ref{fig:mixed}, but for the pure forces
evaluated using equation (\ref{final pure force}).}
\label{fig:pure}   
\end{figure}

Results using equation (\ref{final pure force}) are plotted in figure
\ref{fig:pure}.  We refer to this as the expression for the pure
forces, although it also contains a term evaluated with the
variational distribution.  The total pure forces $F_{\rm pure}^{\rm
  tot}$ on the H and Si atoms sum almost to zero over the range of
bond lengths shown, with an error corresponding to an error in the
bond length of less than 0.003\,\AA, which is about three times the
statistical error.  The equilibrium bond lengths from the zero force
condition are also within statistical errors of the equilibrium bond
length obtained from the minimum in the DMC energy.  The nodal terms
in equation (\ref{final pure force}) are by no means negligible.  The
pure Hellmann-Feynman forces $F_{\rm pure}^{\rm HFT}$ on the H and Si
atoms do not sum to zero, and the force on the H atom is zero at a
bond length which is 0.008(1)\,\AA\ too long, while the force on the
Si atom is zero at a bond length which is 0.010(2)\,\AA\ too long.
The total pure forces are much more accurate than the total mixed
forces, which suggests that the approximate forms of equation
(\ref{final pure force}) and ${\rm d}\Psi_{\rm T} / {\rm d}\lambda$
give rather accurate results in our simulations.  The bond lengths
obtained from the total forces and from the HFT term, and using forces
on the Si and H atoms, within mixed and pure DMC, are summarised in
Table \ref{tab:bond lengths}.  It is clear that the total pure forces
give equilibrium bond lengths differing from that obtained from the
DMC energies by much less than the typical error in DFT bond lengths
of $\sim$0.01\,\AA.

\begin{table}[!pb]
\centering
\caption[]{Equilibrium bond lengths $a$ in \AA\ for the SiH molecule
obtained from the total forces and from the HFT term and using forces
on the Si and H atoms, within mixed and pure DMC.  The equilibrium
bond length obtained from the minimum in the DMC energy curve is
1.5242(6)\,\AA.}
\label{tab:bond lengths}   
\renewcommand{\arraystretch}{1.2}
\setlength\tabcolsep{5pt}
\begin{tabular}{@{}llll@{}}
\hline\noalign{\smallskip}
 & mixed DMC & pure DMC &  \\
\hline\noalign{\smallskip}
$a^{\rm tot}$ (Si) & 1.5140(3) & 1.5259(8)  & \\
$a^{\rm HFT}$ (Si) & 1.5044(8) & 1.5343(16) & \\
$a^{\rm tot}$ (H)  & 1.5150(3) & 1.5227(7)  & \\
$a^{\rm HFT}$ (H)  & 1.5272(9) & 1.5320(12) & \\
\hline
\end{tabular}
\end{table}

\section{Statistical properties of the force estimators}
\label{sec:statistical properties}

A statistical estimate of a quantity should be accompanied by a
confidence interval which indicates the reliability of the estimate.
Consider a probability distribution function (PDF) $P(x)$ which has a
mean value $\mu$ and variance $\sigma^2$.  The mean value of a random
variable $x$ drawn from $P(x)$ may be estimated using the sample mean,
\begin{eqnarray}
\label{mean}
\bar{x}_{N}=\frac{1}{N}\sum_{i=1}^{N} x_{i} \;,
\end{eqnarray}
and the variance estimated using the sample variance,
\begin{eqnarray}
\label{variance}
\sigma_N^2 =\frac{1}{N-1}\sum_{i=1}^{N} \left(x_{i} - \bar{x}_{N}
\right)^2 \;.
\end{eqnarray}
We then say that $\bar{x}_{N}$ lies within the interval
$\left[\bar{x}_{N}-{\sigma_{N}}/\sqrt{N},\,\bar{x}_{N}+{\sigma_{N}}/\sqrt{N}\right]$
with a confidence of 68\,\%.


This statement relies upon the probability distribution satisfying
certain conditions which are the content of the Law of Large Numbers
(LLN) and the Central Limit Theorem (CLT).  The LLN concerns the
statistical convergence
\footnote{By this we mean that the convergence is ``almost sure'' in
the sense that the underlying probability density function approaches
a delta function in the limit of a large sample size.} of the sum in
equation (\ref{mean}) to the true mean value with increasing $N$.
Similarly, the CLT concerns the statistical convergence with
increasing $N$ of the probability distribution from which
$\bar{x}_{N}$ is drawn to a Normal distribution with mean $\mu = \int
xP(x)\, dx$ and variance $\sigma^2/N = 1/N\int (x-\mu)^2 P(x)\, dx$.
For the LLN to be valid it is sufficient that the first moment of
$P(x)$ exists, which requires that $P(x)$ decays faster than
$|x-x_0|^{-2}$, where $x_0$ is a constant.  Similarly, the CLT is
valid if the second moment of $P(x)$ exists, which requires that
$P(x)$ decays faster than $|x-x_0|^{-3}$.

Equations (\ref{eq:variational_energy_3}) and
(\ref{eq:diffusion_energy 2}) show that the VMC and DMC energies are
evaluated as averages of local energies $E_{\rm L}({\bf R}_i)$.  The
probability distribution of the local energies $P(E_{\rm L})$ is
generally non-Gaussian and has ``fat tails''.  The asymptotic
behaviour of $P(E_{\rm L})$ is determined by the singularities in
$E_{\rm L}$.  Trail \cite{trail_2008a} considered singularities in
$E_{\rm L}$ arising from a Coulomb energy divergence at particle
coalescences, divergences in the local kinetic energy at the nodal
surface of $\Psi_{\rm T}$, and in finite systems where $\Psi_{\rm T}$
has the wrong asymptotic behaviour.  The Coulomb energy divergence at
particle coalescences can be removed by imposing the Kato cusp
conditions on $\Psi_{\rm T}$, as is normally done in QMC calculations,
but it is not clear how to remove the divergences at the nodal
surface.  Trail \cite{trail_2008a} showed that the singularities in
$E_{\rm L}$ lead to tails in $P(E_{\rm L})$ which decay as $|E_{\rm
L}-E_0|^{-4}$, where $E_0$ is a constant.  Consequently the energy
estimate is drawn from a Normal distribution whose variance we may
estimate (in the limit of large $N$).  However, the estimated variance
of the local energy is drawn from a distribution of infinite variance
that is not Normal, hence it is not entirely clear how to obtain a
meaningful confidence interval for this estimate.


A similar analysis can be performed for the forces.  Each term in the
force can be written as a sum of local contributions,
\begin{eqnarray}
\label{eq:force sum}
F \simeq \frac{1}{N} \sum_{i=1}^N F_{\rm L}({\bf R}_i) \;.
\end{eqnarray}
For the Hellmann-Feynman term we distinguish between three cases: an
all-electron atom, a smooth local pseudopotential, and a smooth
non-local pseudopotential.  The probability distribution of the
Hellmann-Feynman term for an all-electron atom has slowly decaying
tails of the form $|F_{\rm L}-F_0|^{-5/2}$, where $F_0$ is a constant.
In this case the LLN applies but the CLT is not satisfied, and the
variance of the Hellmann-Feynman force is infinite.  Although a sample
variance can be calculated it does not statistically converge to a
constant value with increasing $N$, and it is not related to the width
of the distribution from which the estimated force is drawn.  The
infinite variance in the Hellmann-Feynman force for an all-electron
atom has long been recognised and methods for dealing with it have
been devised.  Assaraf and Caffarel \cite{assaraf_1999,assaraf_2003}
suggested adding a term to the HFT force which has zero mean but may
greatly reduce the sample variance. (This amounts to adding the term
containing ${\rm d}\Psi_{\rm T} / {\rm d}\lambda$ in the first term on
the right hand side of equations (\ref{eq:vmc force}) and (\ref{final
pure force}).)  A similar approach was developed by Per \textit{et al}
\cite{per_2008}.  Chiesa \textit{et al} \cite{chiesa_2005} developed a
method to filter out the component of the electronic charge density
responsible for the infinite variance.  Using smooth local or
non-local pseudopotentials without singularities at the nucleus also
eliminates the infinite-variance problem arising from the
electron-nucleus interaction.  The terms in the force estimators which
contain ${\rm d}\Psi_{\rm T} / {\rm d}\lambda$ also give probability
distributions with $|F_{\rm L}-F_0|^{-5/2}$ tails, and hence the CLT
is not applicable.  
The origin of the problem is that although $\Psi_{\rm T}$ is zero on
the nodal surface, ${\rm d}\Psi_{\rm T} / {\rm d}\lambda$ is normally
finite.  Trail \cite{trail_2008b} and Attaccalite and Sorella
\cite{attaccalite_2008} have devised rigorous methods within VMC to
eliminate the infinite variance by altering the probability
distribution near the nodes.

\begin{figure}[h!]
\centering
\includegraphics*[width=.7\textwidth]{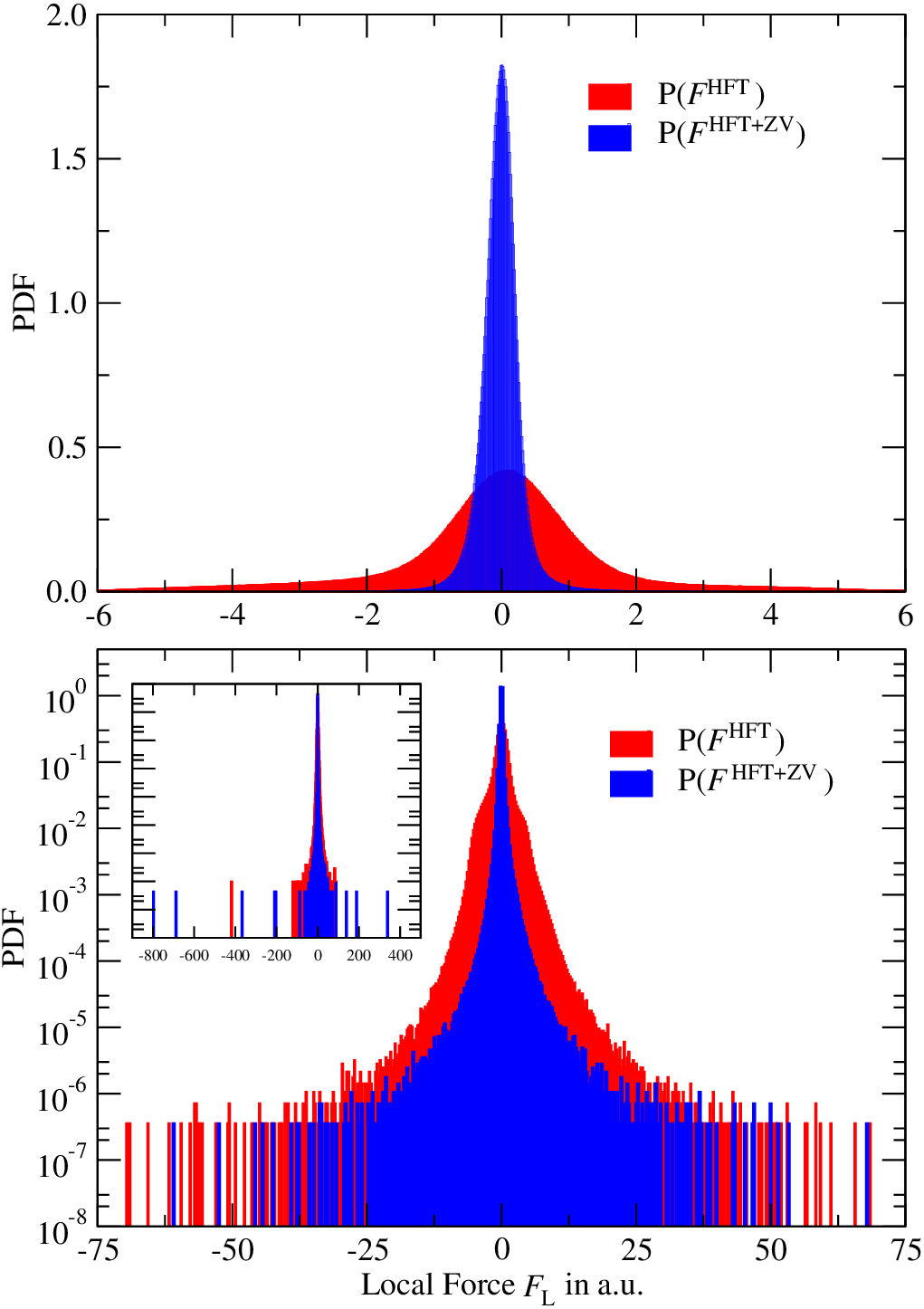}
\caption[]{The probability distribution of the VMC local
Hellmann-Feynman forces (red) and the local Hellmann-Feynman forces
with the zero variance force (blue).  The forces on the Ge atom of a
GeH molecule are shown calculated using non-local pseudopotentials and
a Slater-Jastrow wave function.  The distributions were generated from
10$^7$ samples.  The upper and lower graphs show the same data on
different scales.}
\label{probability distribution forces}   
\end{figure}

Figure \ref{probability distribution forces} shows probability
distributions for VMC local forces on the Ge atom in a GeH molecule.
The upper graph shows that the probability distribution for the sum of
the local Hellmann-Feynman force (the term in equation (\ref{eq:vmc
force}) containing ${\rm d}\hat{H} / {\rm d} \lambda$) and the local
zero-variance force (the term in equation (\ref{eq:vmc force})
containing $\hat{H}$) is much more strongly peaked in the region of
the mean value than the local Hellmann-Feynman force itself.  This
illustrates the operation of the zero-variance term.  The lower graphs
show the same data plotted on different scales.  The inset graph shows
a very wide scale of local forces, and the local zero-variance force
actually has more outlying contributions than the local
Hellmann-Feynman force.  This illustrates the result that the
probability distribution of the local Hellmann-Feynman force has
$|F_{\rm L}-F_0|^{-4}$ tails (for a smooth pseudopotential) and the
CLT holds, while the probability distribution of the local
zero-variance force has $|F_{\rm L}-F_0|^{-5/2}$ tails so that the CLT
is not satisfied.  This seems somewhat counterintuitive; without the
zero-variance term the variance is finite but does not approach zero
when $(\Psi_{\rm T},{\rm d}\Psi_{\rm T}/{\rm d}\lambda)$ are exact
whereas, if we include the zero-variance term, the variance is zero
for the exact case, but infinite for the approximate case.  Analogous
results apply for the DMC forces obtained with the mixed and pure
distributions.

Currently no rigorous procedure has been demonstrated for removing the
infinite variance which arises in DMC calculations of the Pulay
forces.  Although this state of affairs is unsatisfactory, the
probability distribution of the local forces is strongly peaked in the
region of the true value, which suggests that reasonable procedures
for estimating a confidence interval can be devised.

\section{Conclusions}
\label{sec:conclusions}

We have formulated DMC forces evaluated with the pure distribution.
The force can be written as the sum of a Hellmann-Feynman force and a
term which can be cast as an integral over the nodal surface.  The
nodal term can be recast as an average over the pure distribution of a
quantity which depends only on $\Psi_{\rm T}$ and ${\rm d}\Psi_{\rm
  T}/{\rm d}\lambda$.  It is obvious that such an exact expression
must exist, as $\Psi_{\rm T}$ and ${\rm d}\Psi_{\rm T}/{\rm d}\lambda$
fix both the nodal surface and its first order variation, which in
turn fix the DMC energy and its first derivative.  It is however, very
awkward to evaluate an integral over the nodal surface, and therefore
we have adopted an approximate approach to evaluating it which has an
error of order $(\Psi_{\rm T}-\Phi)^2$.  Our final expression
(\ref{final pure force}) consists of integrals over the pure and
variational distributions, both of which exhibit a zero variance
property.

The forces evaluated using equation (\ref{final pure force}) yield
bond lengths for the SiH molecule with an error (with respect to DMC
energy calculations) of much less than 0.01\,\AA, which is acceptable
for most purposes.  Of course, before claiming that the problem of
calculating DMC forces is ``solved'', accurate results for heavier
atoms and larger systems are required.  To this we should add the
issues of efficient generation of the pure probability distribution
for large systems, overcoming the ``infinite variance'' described in
Section \ref{sec:statistical properties}, obtaining accurate forms for
$\Psi_{\rm T}$ and ${\rm d}\Psi_{\rm T}/{\rm d}\lambda$, and
(hopefully) removing the second approximation in equation (\ref{final
pure force}).  Notwithstanding these caveats, we believe that
substantial progress has been made in calculating DMC forces, and that
their evaluation will become routine in the not-too-distant future.

\section{Acknowledgments}
\label{sec:acknowledgments}

This work has been supported by the Engineering and Physical Sciences
Research Council (EPSRC), UK\@.  PDH was supported by a Royal Society
University Research Fellowship.  Computing resources were provided by
the Cambridge High Performance Computing Service.

\end{document}